\documentclass{article}

\usepackage{arxiv}

\usepackage[utf8]{inputenc} 
\usepackage[T1]{fontenc}    
\usepackage{hyperref}       
\usepackage{url}            
\usepackage{booktabs}       
\usepackage{amsfonts}       
\usepackage{nicefrac}       
\usepackage{microtype}      
\usepackage{lipsum}		
\usepackage{graphicx}
\usepackage{natbib}
\usepackage{doi}
\usepackage{amsmath}
\usepackage{subcaption}
\usepackage{arydshln}
\usepackage{booktabs}
\usepackage{xcolor}

\hypersetup{
    colorlinks=true,
    citecolor = blue,
    linkcolor=blue,
    filecolor=magenta,      
    urlcolor=blue,
    pdftitle={The Addams family},
    pdfpagemode=FullScreen,
    }

\newcommand{\new}[1]{{\color{black}{#1}}}
\newcommand{\newnew}[1]{{\color{black}{#1}}}
\newcommand{\newnewnew}[1]{{\color{black}{#1}}}
\newcommand{\newnewnewnew}[1]{{\color{black}{#1}}}

\title{On the Addams family of discrete frailty distributions for modelling multivariate case I interval-censored data}

\author{ Maximilian Bardo \\
	Department of Medical Statistics\\
	University Medical Center G\"{o}ttingen\\
	G\"{o}ttingen, Germany \\
	{\href{email:maximilian.bardo@proton.me}{maximilian.bardo@proton.me}} \\
	\And
	Niel Hens \\
	I-BioStat, Data Science Institute\\
	  Hasselt University\\
	Hasselt, Belgium \\
 Centre for Health Economics Research and \\ Modelling Infectious Diseases, \\Centre for the Evaluation of Vaccination,\\ Vaccine \& Infectious Disease Institute \\
    University of Antwerp\\
    Wilrijk, Belgium \\
 \And
	Steffen Unkel \\
	Department of Medical Statistics\\
	University Medical Center G\"{o}ttingen\\
	G\"{o}ttingen, Germany \\
    Faculty V: School of Life Sciences \\
    University of Siegen \\
    Siegen, Germany\\
}

\renewcommand{\shorttitle}{On the Addams family of discrete frailty distributions}

\hypersetup{
pdftitle={A template for the arxiv style},
pdfsubject={q-bio.NC, q-bio.QM},
pdfauthor={David S.~Hippocampus, Elias D.~Striatum},
pdfkeywords={First keyword, Second keyword, More},
}

\begin{document}
\maketitle

\begin{abstract}
Random effect models for time-to-event data, also known as frailty models, provide
a conceptually appealing way of quantifying association between
survival times and of representing heterogeneities resulting from factors which
may be difficult or impossible to measure.
In the literature, the random effect is usually assumed to have a continuous distribution. 
However, in some areas of application, discrete frailty distributions may be more appropriate.
The present paper is about the implementation and interpretation of the Addams family of discrete frailty distributions. 
We propose methods of estimation for this family of densities in the context of shared frailty models for the hazard rates for case I interval-censored data. 
Our optimization framework allows for stratification of random effect distributions by covariates.
We highlight interpretational advantages of the Addams family of discrete frailty distributions and the $K$-point distribution as compared to other frailty distributions.
A unique feature of the Addams family and the $K$-point distribution is that the support of the frailty distribution depends on its parameters.
This feature is best exploited by imposing a model on the distributional parameters, resulting in a model with non-homogeneous covariate effects that can be analysed using standard measures such as the hazard ratio.
Our methods are illustrated with applications to multivariate case I interval-censored infection data.
\end{abstract}


\section{Introduction}

Multivariate time-to-event data are commonly encountered in the life sciences.
An example is the time to occurrence of multiple non-lethal events within the same individual. 
In this setting, the individual can be thought of as forming a cluster in which event times are likely to be correlated.
An alternative point of view is that there is heterogeneity across individuals (or clusters) due to characteristics that may be difficult or impossible to measure.
Random effect (RE) models for time-to-event data, also known as frailty models, offer a conceptually appealing approach to quantify these associations within clusters and to model unobserved heterogeneity across individuals (or clusters) \citep{Duchateau.2008,Hougaard.2000,Wienke.2010}.

While the majority of the existing literature assumes continuous frailty distributions, such as the gamma ($\mathcal{G}$), inverse Gaussian, or log-normal, for some applications discrete frailty distributions may be more appropriate. 
One such application is the study of infectious diseases transmitted via close contact, where unobserved heterogeneity with respect to becoming infected with pathogens of interest may be represented and analysed by latent risk groups.

This paper discusses the interpretation of discrete frailty models with parameter dependent support. 
We propose interpreting the discrete frailties as ordered latent risk categories.
We extend this analysis by allowing the parameters of the discrete frailty distribution to depend on covariates. 
The ordered nature of discrete frailty facilitates comparing latent risk categories within or across different strata (as defined by covariate values) using hazard ratios, along with probabilities of belonging to specific risk categories. 
This approach facilitates the analysis of non-homogeneous populations or non-homogeneous covariate effects, which might lead to a deeper understanding of the problem under study.

In this context we discuss the ``Addams'' family ($\mathcal{AF}$) of discrete frailty distributions introduced by \cite{Farrington.2012}. 
This family is homogeneous in the sense that it consists of discrete distributions, except for the continuous $\mathcal{G}$ distribution, 
which serves as a reference case of constant association within clusters over time. 
The discrete distributions within the $\mathcal{AF}$ are scaled variants of negative binomial, shifted negative binomial, binomial, and Poisson distributions. 
For this family, we investigate the hazard ratios conditional on risk group membership within or across different strata. 
Furthermore, we introduce estimation routines for the $\mathcal{AF}$ in the context of case I interval-censored data \citep{Sun.2007}. 
Our optimization framework allows for stratification of frailty distributions along covariates.
We apply the $\mathcal{AF}$ to data from a serological survey of human papillomaviruses \citep{Mollema.2009,Scherpenisse.2012}.

The structure of the paper is as follows: Section \ref{LitRev} presents the time-invariant shared frailty model and reviews the literature on discrete frailty models, comparing the $\mathcal{AF}$ to other discrete frailty distributions. We further discuss the interpretation of discrete frailty models for which the support is parameter dependent. Section \ref{Frailty} examines the conditional time-to-event model resulting from the $\mathcal{AF}$ of distributions. Section \ref{Likelihood} outlines the estimation framework and optimization algorithm. In Section \ref{app}, the $\mathcal{AF}$ is applied to multivariate serological survey data. Finally, Section \ref{conc} offers concluding remarks. 

\section{Discrete Frailty Models}\label{LitRev}

Let $i$ refer to a cluster, $i = 1, \dots, n$, and let $n$ be the number of clusters observed. 
Each cluster contains $J$ units, and $T_i^{(j)}>0$ denotes the time-to-event random variable (RV) for the $j ^{\text{th}}$ unit, $j=1,\dots,J$, within the $i ^{\text{th}}$ cluster. 
All indices are unique, such that $i\neq i'$ and $j\neq j'$.
The random vectors $\boldsymbol{T}_i=[T_i^{(1)},\dots, T_i^{(J)}]^T$ and $\boldsymbol{T}_{i'}$ are independent given the covariates $\boldsymbol{x}_i$ and $\boldsymbol{x}_{i'}$. 
Within a cluster, $T_i^{(j)}$ and $T_i^{(j')}$ are independent given the random and unobservable time-invariant cluster-specific frailty $Z_i$ and covariates. 
The non-negative RV $Z_i$ has density or probability mass function $g(z|\tilde{\boldsymbol{x}}_i) \equiv g_i(z)$, where the index $i$ is relevant only if the distribution or its parameters depend on (a subset of) covariates denoted by $\tilde{\boldsymbol{x}}_i$. 
\newnewnew{Technically, $\tilde{\boldsymbol{x}}_i$ could include unit-specific covariates. 
However, as $Z_i$ is shared within the cluster $i$, $\tilde{\boldsymbol{x}}_i$ will typically also be shared within the cluster $i$, without containing covariates that differ across the units. }
\newnewnewnew{
For unit-specific covariates, correlated frailties \citep{Hens.2009} may be more appropriate, where the correlation parameter between the frailties of a cluster may depend on cluster-invariant covariates, while unit-specific frailty parameters may depend on unit-specific covariates. }
Given covariates, $Z_i$ and $Z_{i'}$ are independent. 
We do not distinguish in language between the random $Z_i$ and a realisation $z_i$ and refer to both as frailty or RE.

The conditional hazard rate, is assumed to be of the form
\begin{align}
\lambda^{(j)}_{i}(t|Z_i) &= Z_i \exp\{{\boldsymbol{x}_i^{(j)}}^T \boldsymbol{\beta}^{(j)}\} \lambda^{(j)}_0(t), \label{model}
\end{align}
with cluster- and unit-specific covariate vector $\boldsymbol{x}_i^{(j)}$, and parameters $\boldsymbol{\beta}^{(j)}$, as well as unit-specific baseline hazard rate $\lambda^{(j)}_0(t)$.
Note that for brevity, the dependence of quantities, such as hazard rates or densities, on $\boldsymbol{x}_i^{(j)}$ or $\tilde{\boldsymbol{x}}_i$ is indicated by a superscript $(j)$ or subscript $i$, respectively.

In much of the existing literature, $Z_i$ is typically considered a continuous random variable, with common choices being the log-normal or $\mathcal{G}$ distributions. 
Nonetheless, discrete frailty distributions have also been explored for both univariate and multivariate data. 
\new{A prominent choice is the $K$-point distribution \citep{Palloni.2017,Bijwaard.2014,Begun.2000,Gasperoni.2020,Pickles.1994,Choi.2012,Choi.2014,Tronco.2018}, i.e. a frailty distribution with (ordered) support parameters $\{z_{(1)}, \dots, z_{(K)}\} \in \mathbb{R}^K_{\geq 0}$ and corresponding probability parameters $\text{pr}_m = g_i(z_{(m)})>0$ for $m =1,\dots,K$, and $\sum_{m=1}^K \text{pr}_m = 1$ \citep{Wienke.2010}. 
Other choices are the binomial ($\mathcal{B}$) \citep{Ata.2013}, negative binomial ($\mathcal{NB}$) \citep{Ata.2013,Caroni.2010}, geometric \citep{Caroni.2010,Cancho.2021,Choi.2012,Choi.2014}, Poisson ($\mathcal{P}$) \citep{Ata.2013,Caroni.2010,Cancho.2020b,Choi.2012,Choi.2014}, and the hyper-Poisson distributions \citep{Mohseni.2020,Souza.2017}. 
The framework of the zero-inflated and zero-modified power series (ZMPS) distributions, where the probability of $Z=0$ is modified by an additional parameter as compared to the discrete reference distribution, has been investigated by \cite{Cancho.2018,Cancho.2020} and \cite{Molina.2021}, respectively. In particular, the zero-inflated geometric, $\mathcal{P}$ and logarithmic distributions and the zero-modified geometric and $\mathcal{P}$ distributions are investigated. 
The Addams family ($\mathcal{AF}$), as conceptualised by \cite{Farrington.2012}, includes (shifted and) scaled $\mathcal{NB}$, as well as scaled $\mathcal{B}$ and $\mathcal{P}$ distributions, and the $\mathcal{G}$ distribution as a continuous special case.}

\new{Modelling a whole family of distributions, such as the $\mathcal{AF}$, is preferable to modelling a given (discrete) distribution, such as the $\mathcal{P}$, since the latter strategy typically severely limits the patterns of heterogeneity as measured by the relative frailty variance ($\operatorname{RFV}(\boldsymbol{t})=\frac{\operatorname{Var}(Z|\boldsymbol{T}>\boldsymbol{t})}{\operatorname{E}(Z|\boldsymbol{T}>\boldsymbol{t})^2}$, $\boldsymbol{t}=[t^{(1)},\dots,t^{(J)}]^T$) to either monotonically increasing or monotonically decreasing trajectories over time; for examples see \cite{Farrington.2012} and \cite{Bardo.2023a}. 
It can be shown that the long-term trajectory of the RFV (or association, as measured by the cross-ratio function [$\operatorname{CRF}(\boldsymbol{t})= 1 + \operatorname{RFV}(\boldsymbol{t})$]) is determined with positive probability by the smallest value of the frailty distribution ($z_{(1)}$). 
Specifically, if $z_{(1)}=0$, $\operatorname{RFV}$ and $\operatorname{CRF}$ approach infinity as time approaches infinity. 
Conversely, if $z_{(1)}>0$, the RFV approaches zero and the CRF approaches one. 
Therefore, the choice of a discrete frailty distribution has an enormous impact on the model, even if $g_i(0)>0$ or $z_{(1)}>0$ is very small, and even more so if the model's trajectory of RFV is monotone. 
The $\mathcal{AF}$ achieves greater flexibility in the trajectory of heterogeneity and association by incorporating discrete frailty distributions for which $z_{(1)} = 0$ and frailty distributions for which $z_{(1)}>0$, and is thus able to induce increasing and decreasing trajectories of the RFV (CRF). 
Thus, in the case of $\mathcal{AF}$, the trajectory of the RFV can be informed by the data in the fitting process.  
This is a rare property among frailty distributions, and for the discrete distributions mentioned above it is only possible for the $K$-point distribution and the ZMPS. However, in these cases the decision between a decreasing or increasing long-term trajectory lies on the edge of the parameter space \citep{Bardo.2023a}, which is not the case for the $\mathcal{AF}$. 
As optimising on the edge of the parameter space is usually difficult, this is an advantage when using the $\mathcal{AF}$. 
Note, however, that the $K$-point distribution and the ZMPS are able to induce non-monotone trajectories of the RFV (CRF), which is not the case for the $\mathcal{AF}$. 
We will discuss the $\mathcal{AF}$ and its special cases in Section \ref{Frailty}.}

\new{From an interpretive point of view, the $\mathcal{AF}$ and $K$-point distributions offer a unique perspective because the support is also subject to estimation.
The support of other discrete frailty distributions is usually the natural number including zero ($\mathbb{N}_{\geq 0}$).
Therefore, the interpretation of discrete frailty models often focuses on the cure rate, i.e. those who are not susceptible to the event of interest (see e.g. \cite{Souza.2017, Cancho.2018, Cancho.2020, Cancho.2020b, Mohseni.2020, Cancho.2021, Molina.2021}).
Subject-related interpretations are also common. 
\cite{Caroni.2010} suggest interpreting discrete frailties as the unobservable number of flaws in a unit or exposure to damage on an unknown number of occasions.
Similar interpretations can be found in \cite{Ata.2013} for time-to-event models on earthquake data.
Both studies consider discrete frailty distributions with support on $\mathbb{N}_{\geq 0}$.
However, due to the latent nature of the frailty, such concrete interpretations are difficult because the hazard ratio (HR) of, say, two events versus one event would be fixed by the model structure at $\text{HR}=2$ if the support is fixed at $\mathbb{N}_{\geq 0}$.
However, if the true HR is less than two, the probability mass of frailty could be shifted to the right relative to the distribution of the number of hits.
Consequently, a more abstract interpretation such as the ``effective'' number of harms would be more appropriate.
The $K$-point frailty distributions are often interpreted as representing sub-populations, such as unobservable carriers of certain disease genotypes (e.g., \cite{Pickles.1994, Begun.2000, Wienke.2010, Bijwaard.2014}). 
\cite{Palloni.2017} interpret a delayed binary frailty as the effect of adverse early life conditions on adult mortality. 
Using the $K$-point distribution, where $K$ is also subject to estimation, \cite{Gasperoni.2020} interpret the discrete frailties as (an unknown number of) latent sub-populations. 
They suggest calculating HRs between these sub-populations, which is interpretable as the support being subject to estimation.}

In the present paper, we endorse the interpretation of discrete frailties as latent sub-populations and expand upon it.
Note that we focus on discrete frailty distributions for which the support $\Omega$ is parameter dependent, which is mainly the case for the $K$-point distribution and, \new{although more restricted}, the $\mathcal{AF}$, as will be seen in Section \ref{Frailty}.
Let $\Omega_i = \{z_{i,(1)}, z_{i,(2)},\dots\}$, with $z_{i,(1)}\geq 0$ and $z_{i,(k)}<z_{i,(k+1)}$, represent the support of discrete RV $Z_i$, where the distribution parameters might depend on $\tilde{\boldsymbol{x}}_i$. 
We define that $\tilde{\boldsymbol{x}}_i$ constitutes a stratum of the population. 
We also consider $z_{i,(k)}$ as the conditional hazard-determining value for an individual in the $k^{\text{th}}$ risk category (RC) within the stratum which is defined by $\tilde{\boldsymbol{x}}_i$.

For discrete frailty distributions for which the support depends on parameters, the within-stratum hazard ratio, $\text{HR}_\text{W}(k)=\frac{\lambda_i^{(j)}(t|Z_i=z_{i,(k+1)})}{\lambda_i^{(j)}(t|Z_i=z_{i,(k)})}=\frac{z_{i,(k+1)}}{z_{i,(k)}}$, might be analysed.
The $\text{HR}_\text{W}(k)$ compares the hazard of the $k^{\text{th}}$ and ${(k+1)}^{\text{th}}$ RC of stratum $i$. 
This is in line with \cite{Gasperoni.2020}, except for the presence of different strata for the distribution of the frailty, i.e. the $\text{HR}_\text{W}$ might differ for $\tilde{\boldsymbol{x}}_i$ and $\tilde{\boldsymbol{x}}_{i'}$.
Moreover, due to the presence of different strata for the frailty distribution, an across-stratum analysis can be conducted by computing the across-stratum hazard ratio,  $\text{HR}_\text{A}(k)=\frac{\lambda_i^{(j)}(t|Z_i=z_{i,(k)})}{\lambda_{i'}^{(j)}(t|Z_{i'}=z_{i',(k)})}=\frac{z_{i,(k)}}{z_{i',(k)}}$, for $\tilde{\boldsymbol{x}}_i \neq \tilde{\boldsymbol{x}}_{i'}$, and equality in the remaining covariates.
{If the support between stratum $i$ and $i'$ is very different, it might be more desirable to analyse $\text{HR}_\text{A}(k;k')=\frac{z_{i,(k)}}{z_{i',(k')}}$ for all $k'$ for which $P(Z_{i'} \leq z_{i',(k')}) \in [P(Z_i \leq z_{i,(k-1)}), P(Z_i \leq z_{i,(k)}))$  or the closest quantiles of $Z_i, Z_{i'}$ if no such $k'$ exists.
In order to put the analysis of the $\text{HR}_\text{W}$ and $\text{HR}_\text{A}$ into context, they should always be accompanied by reporting the distribution of the RCs.
This allows for a separate but accompanying analysis of the distribution of individual heterogeneity via the distribution of RCs and the magnitude related impact of the RCs on the hazard rates.
}


The approach of imposing a model on the distribution parameters of the frailty has some similarity to generalized additive models for location, scale, and shape parameters for the population time-to-event distribution{, i.e. for the time-to-event distribution with the frailty marginalized out.}
However, modelling the determinants of the randomness of hazard rates via covariates might be more intuitive, as the hazard rates are usually the standard approach for modelling time-to-event data.
\newnewnew{This approach is not new per se. 
It can be found, for example, in \cite{Aalen.2008} and is also quite common in the field of discrete frailty modelling \citep{Choi.2012,Choi.2014,Molina.2021,Cancho.2018,Cancho.2020}, and random slopes could also be interpreted in this way. 
What is new, however, is the type of analysis with
within-stratum and across-stratum hazard ratios. 
Note that this type of analysis has no counterpart for continuous frailty distributions, as there is no $k^{\text{th}}$ RC. 
Nor does such an analysis make sense for discrete frailty models where the support of the frailty distribution is set by assumption, e.g. to $\mathbb{N}_{\geq 0}$, as this would fix the $\text{HR}_{\text{W}}$ and $\text{HR}_{\text{A}}$ by assumption. 
Therefore, an analysis via the $\text{HR}_{\text{W}}$ and $\text{HR}_{\text{A}}$ is unique to discrete frailty models, where the support of the frailty distribution varies with (a subset of) distribution parameters that may depend on stratum membership. 
In this case, the data inform the support of the frailty distribution by likelihood criteria making it suitable to represent latent RCs. 
This allows the analysis of non-homogeneous covariate effects using common measures such as HRs as described above and probabilities of belonging to a particular RC. 
} 
This might in particular be helpful in communicating the results of heterogeneous (covariate) effects to an audience outside of statistics such as medical doctors.       
 \label{intro}

\section{The Addams Family of Discrete Frailty Distributions}

Let $\Lambda_0^{(j)}(t)$ denote the cumulative baseline hazard rate $\int_0^{t} \lambda_0^{(j)}(u)du$.
Moreover, $\Lambda_i(\boldsymbol{t}) = \sum_{j=1}^J \exp\{{\boldsymbol{x}_i^{(j)}}^T\boldsymbol{\beta}^{(j)}\} \times$ $\Lambda^{(j)}_0(t^{(j)})$.
We further ignore that the parameters of the frailty distribution may depend on $\tilde{\boldsymbol{x}}_i$ in the first part of this section and come back to this issue in the latter part of this section.

The RFV that induces the $\mathcal{AF}$ equals $\operatorname{RFV}(\boldsymbol{t}) = \gamma \exp\{\alpha \mu  \Lambda_i(\boldsymbol{t})\}$,
with $\gamma, \mu = \operatorname{E}(Z) \in \mathbb{R}_{>0}$, and $\alpha \in \mathbb{R}$.
As shown in \cite{Farrington.2012}, the Laplace transform $\mathcal{L}(s) = \int_0^{\infty} \exp\{-zs\}g(z)dz$ of the $\mathcal{AF}$ equals
\begin{align}
\mathcal{L}(s) = \begin{cases}     \bigg( \big(1 -\frac{\gamma}{\alpha}\big)\exp\{-\alpha \mu  s\} + \frac{\gamma}{\alpha}  \bigg)^{\frac{1}{\alpha - \gamma}} & \text{if } \alpha \neq \gamma, \alpha \neq 0, \\
      \exp\bigg \{ \frac{1}{\gamma} \big ( \exp\{- \gamma \mu s \} - 1 \big) \bigg \} & \text{if  } \alpha = \gamma,\\
      (1 + \gamma \mu s)^{-\frac{1}{\gamma}} & \text{if } \alpha = 0. \end{cases} \label{LaplaceAddams}
\end{align}
Note that $\mu$ may be set to one for identification purposes. 
Hence, the $\mathcal{AF}$ is a two parameter distribution. 
We will keep $\mu$ in the notation, however,  as we will explicitly model the parameter via $\mu(\tilde{\boldsymbol{x}}) = \exp\{\tilde{\boldsymbol{x}}^T \boldsymbol{\beta}^{(0)}\}$, 
where $\boldsymbol{\beta}^{(0)} \in \mathbb{R}$ is an additional set of parameters which is usually interpreted as proportional hazard factor on the baseline hazard rate.

\begin{table}[ht]
\centering
\caption{\newnewnew{RFV and distribution parameters of the Addams family and support.}}  \label{table:ag}
\begin{tabular}[t]{ p{2cm}p{2.5cm}p{6cm}p{3.5cm}  }
\toprule
Parameters & $Z \sim$ & Distribution Parameters & Support\\[1.25ex]
\hline
& & & \\
$\gamma > 0 > \alpha$ & $\psi \mathcal{NB}_{>0}(\nu,\pi)$ & $\nu = \frac{1}{\gamma - \alpha}$ (number of successes), & $\psi \times $ \\ && $\pi = \frac{-\alpha}{\gamma - \alpha}$ (success probability)& $\{\nu, 1+\nu,2+\nu,\dots\}$ \\[1.05ex]
$\gamma > 0 = \alpha$ & $\mathcal{G}(\gamma^{-1},{\gamma^*})$ & $\gamma^{-1}$ (shape), ${\gamma^*} = (\mu \gamma)^{-1}$ (rate)& $\mathbb{R}_{>0}$  \\[1.5ex]
$ \alpha = \gamma > 0$ & $\psi \mathcal{P}({\lambda^*})$ & ${\lambda^*} = \gamma^{-1}$ (rate)& $\psi \times \{0,1,2\dots\}$ \\[1.5ex]
$\gamma > \alpha > 0$ & $\psi\mathcal{NB}(\nu,\pi)$ & $\nu = \frac{1}{\gamma - \alpha}$ (number of successes),& $\psi \times$ \\&& $\pi = \frac{\alpha}{\gamma}$ (success probability)& $\{0,1,2,\dots\}$\\[1.5ex]
$\alpha > \gamma > 0$ & $\psi\mathcal{B}(b,\pi)$ & $b = (\alpha-\gamma)^{-1}$ (number of trials), & $\psi \times$ \\ && $\pi = \frac{\alpha - \gamma}{\alpha}$ (success probability)& $\{0,1,\dots,b\}$\\
\bottomrule
\end{tabular}
\end{table}


The Laplace transform uniquely determines the distribution of $Z$, as shown in Table~\ref{table:ag} \citep{Farrington.2012}. 
\newnewnewnew{The $\mathcal{AF}$ consists of different scaled and possibly shifted discrete distributions. 
The scaling parameter $\psi$ of the corresponding discrete distributions is equal to $\mu {\lvert \alpha \rvert}$,
and the parameter $\alpha$ selects one of the distributions of the $\mathcal{AF}$.  
If $\alpha<0$, the scaled and shifted negative binomial ($\psi \mathcal{NB}_{>0}$) is chosen, where the support is shifted to the right by the parameter $\nu$, resulting in a model without a latent non-susceptible sub-population. 
If $\alpha>0$, a frailty distribution is chosen that includes a non-susceptible latent sub-population. 
For $\gamma > \alpha > 0$, the non-shifted scaled negative binomial ($\psi\mathcal{NB}$) and for $\alpha = \gamma$ the scaled Poisson ($\psi\mathcal{P}$) distribution is selected. 
In the case of $\alpha > \gamma $, the frailty distribution is scaled binomial ($\psi\mathcal{B}$), resulting in a model with an upper bound of the frailty that limits the maximum deviation of RC-conditioned survival curves between the susceptible latent sub-populations and a finite number of RCs including a non-susceptible group. }
Note that for the case $\alpha>\gamma$, $(\alpha - \gamma)^{-1}$ has to be an integer in order for $\mathcal{L}(s)$ to be a valid Laplace transform \citep{Farrington.2012}. 
\newnewnewnew{The continuous exception in the $\mathcal{AF}$ is the $\mathcal{G}$ distribution, which results from $\alpha=0$. }

The parameter $\alpha$ plays a unique role in the context of discrete shared frailty modelling.
For discrete shared frailty models the RFV (CRF) either approaches zero (one) or infinity \citep{Bardo.2023a}.
Therefore, it is desirable to have a continuous exception within a family of discrete shared frailty distributions for which the RFV (CRF) does not approach zero (one) or infinity with time approaching infinity.
Within the context of the $\mathcal{AF}$ this is the $\mathcal{G}$ which arises for $\alpha = 0$.
In that case the RFV (CRF) is constant, a shape that is impossible for a discrete shared frailty model to generate.
The nested structure might be utilized to test for a constant RFV (CRF) within the $\mathcal{AF}$. 

Moreover, the $\mathcal{AF}$ can choose between a monotonically decreasing or increasing trajectory of the RFV (CRF) through the sign of $\alpha$.
This is unprecedented in the context of discrete shared frailty models. 
Though the ZMPS distribution is able to create decreasing trajectories of the RFV (CRF), this involves the edge of the parameter space for its deflation/inflation parameter.
The opposite is true for the $K$-point distribution:
its RFV (CRF) approaches zero (one) in the long run unless $z_{(1)}$ is equal to zero, which again involves the edge of the parameter space.
For the $\mathcal{AF}$, the parameter $\alpha$ chooses between a decreasing or increasing RFV (CRF) by determining the support of the discrete distribution without involving the edge of parameter space. 
If $\alpha > 0$, $z_{(1)} = 0$ and a cure fraction exists. 
This induces an increasing trajectory of the RFV (CRF). 
If $\alpha < 0$ instead, $z_{(1)}>0$ and no cure fraction exists. 
This induces a decreasing trajectory of the RFV (CRF); 
see \cite{Bardo.2023a} for a discussion of the shape of the RFV (CRF) for discrete frailty models.  

\begin{figure}[ht]
\centering
\begin{subfigure}{0.45\textwidth}
    \includegraphics[scale=.265]{{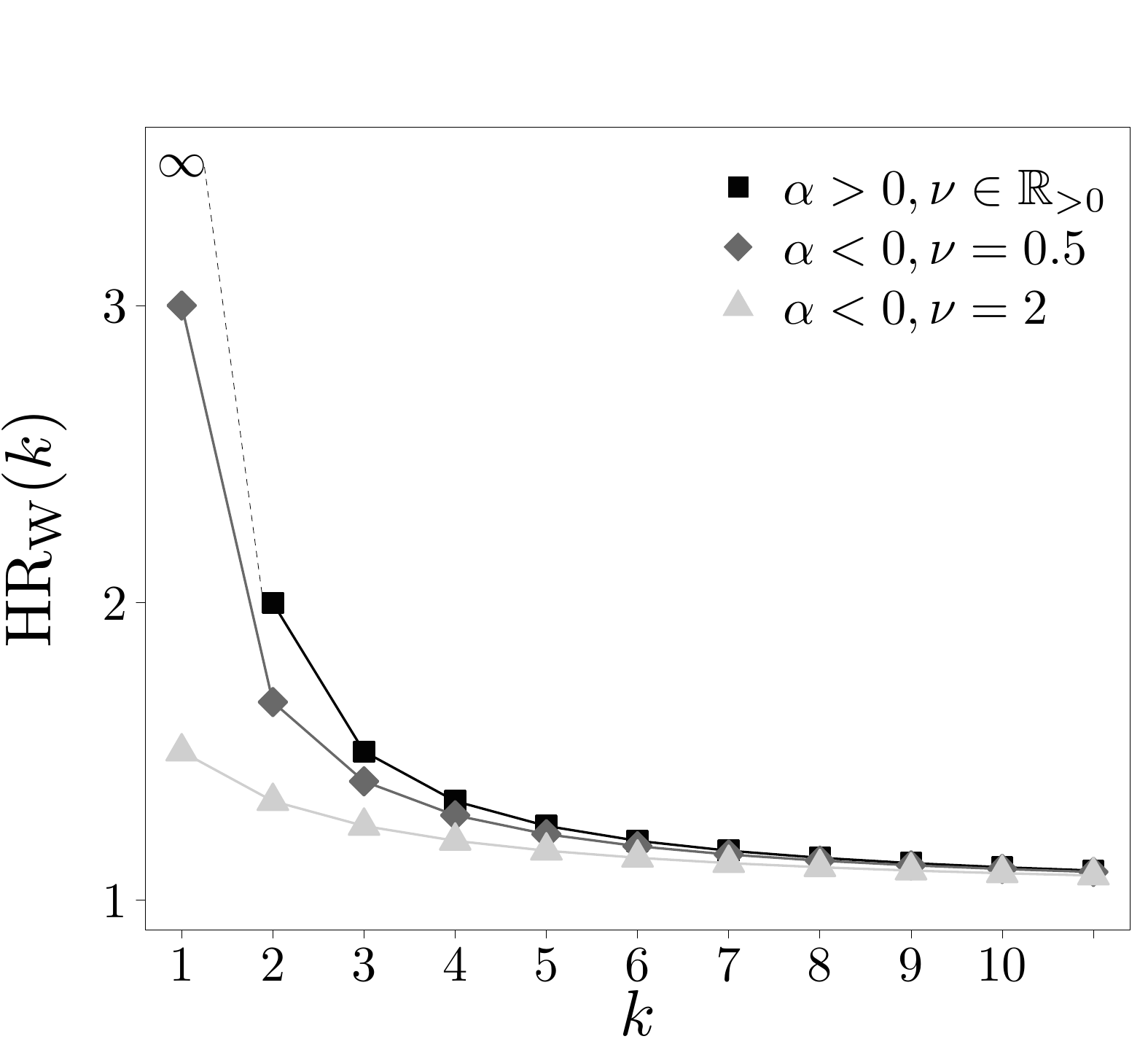}} 
    \subcaption{Within stratum HR for $\alpha>0$ and $\alpha<0$ versus RC.}
    \label{fig:HR_W}
\end{subfigure}
\begin{subfigure}{0.45\textwidth}
    \includegraphics[scale=.265]{{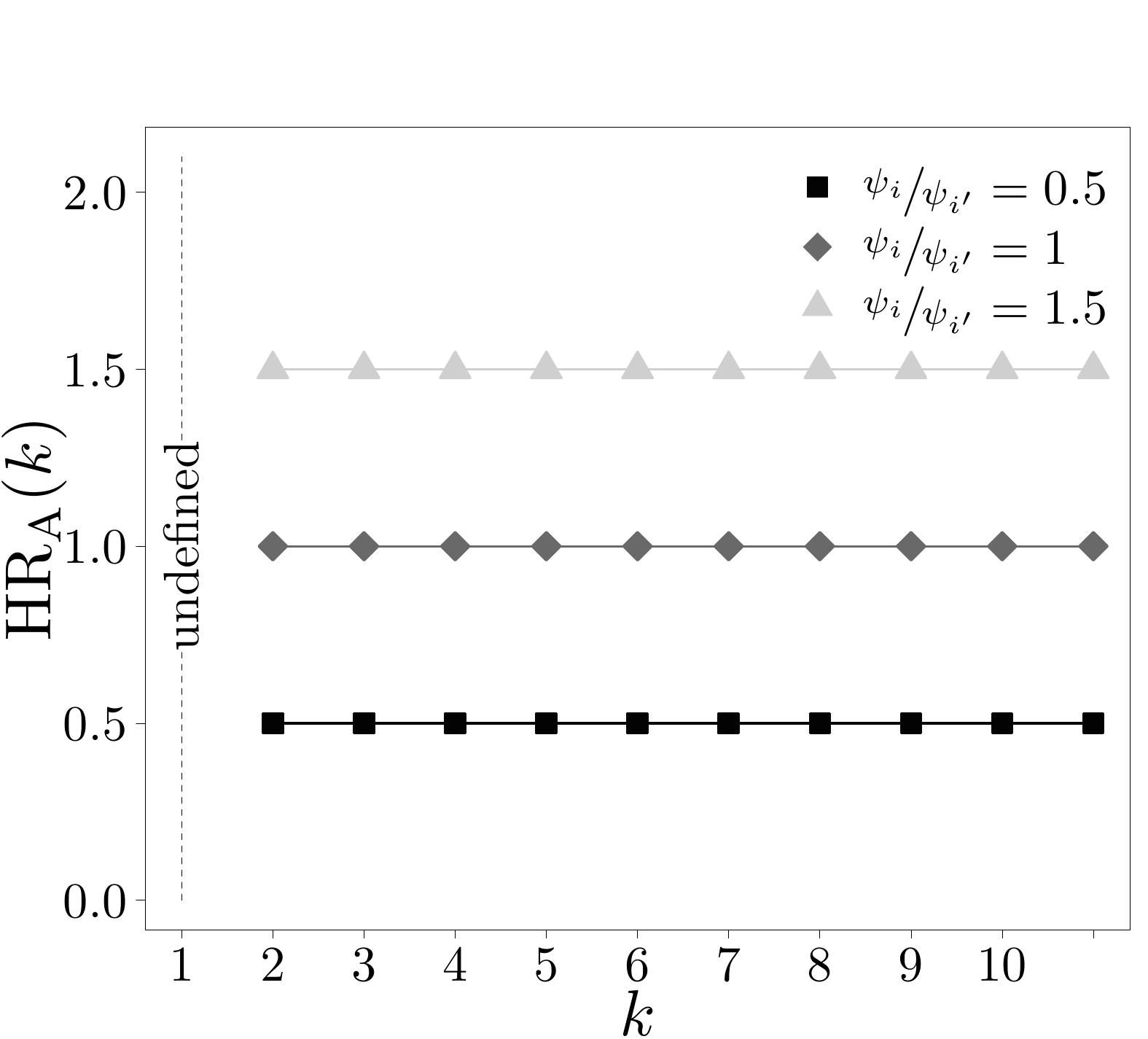}} 
    \subcaption{Across-stratum HR for $\alpha_i,\alpha_{i'}>0$ \new{versus RC $k$}.}
    \label{fig:HR_app}
\end{subfigure}
\begin{subfigure}{0.45\textwidth}
    \includegraphics[scale=.265]{{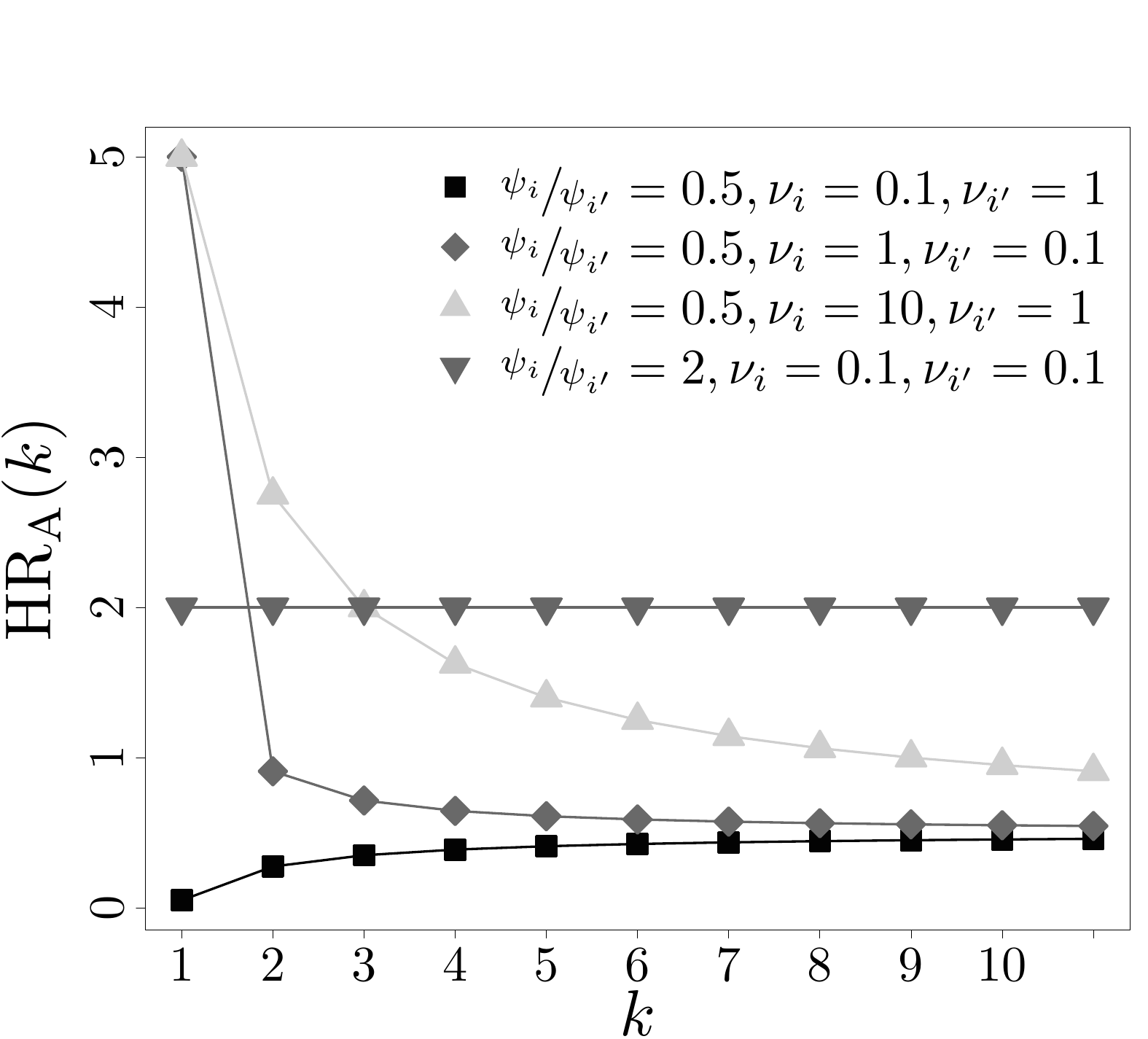}}
    \subcaption{Across-stratum HR for $\alpha_i,\alpha_{i'}<0$ \new{versus RC $k$}.}

    \label{fig:HR_ann}
\end{subfigure}
\begin{subfigure}{0.45\textwidth}
    \includegraphics[scale=.265]{{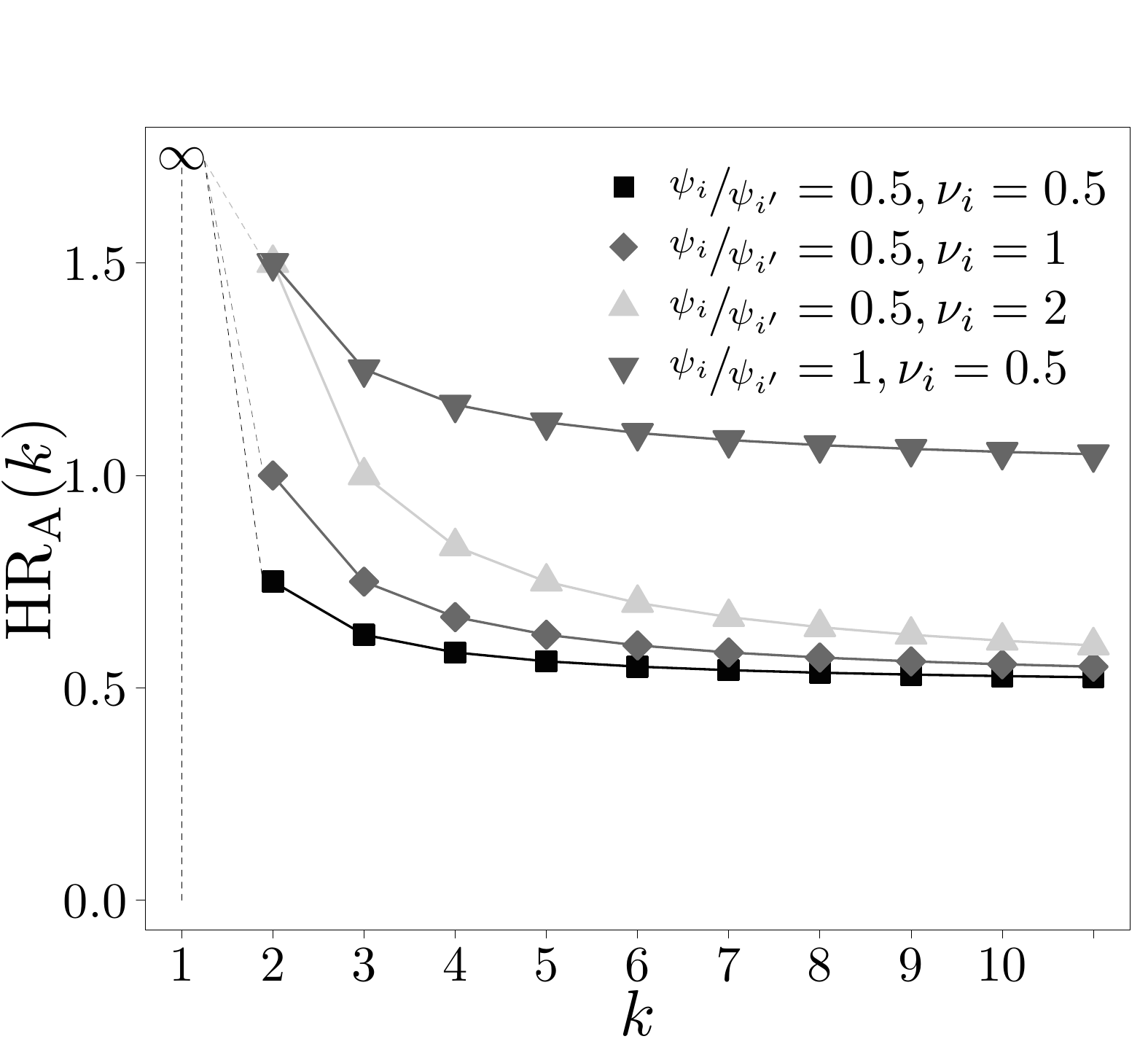}}
    \subcaption{Across-stratum HR for $\alpha_i <0,\alpha_{i'}>0$ \new{versus RC $k$}.}

    \label{fig:HR_anp}
\end{subfigure}
\caption{Within- and across-stratum \new{hazard ratios versus RC $k$ for various members of the Addams family.}} \label{fig:HR_WA}
\end{figure}

The feature of the support being dependent on the distribution parameters offers the possibility of a meaningful interpretation of the $\text{HR}_{\text{W}}$. 
Figure~\ref{fig:HR_WA}\subref{fig:HR_W} shows examples of the $\text{HR}_\text{W}$ for $\alpha>0$ and $\alpha < 0$. 
If $\alpha>0$, $\text{HR}_\text{W}(1) \equiv \infty$ and $\text{HR}_\text{W}(k) = \frac{k}{k-1}$ for $k\geq 2$. 
\newnewnew{Note that there may be an upper bound on the RCs if $\psi\mathcal{B}$ is chosen. 
In this case, the upper bound of the frailty as well as the number of RCs chosen  through the fitting procedure may be the main component of the analysis, e.g. by comparing the RC-related survival curve of the upper bound versus another RC. }
If $\alpha<0$, $\text{HR}_\text{W}(k)=\frac{\nu+k}{\nu +k-1}$ which approaches $\frac{k}{k-1}$ for large $k$.
So there is reasonable flexibility for the first few within-stratum HRs \newnew{and the focus may be on $\text{HR}_{\text{W}}(1)$, where the model is more flexible than for later RCs.
This is less flexible than the $K$-point distribution is, provided that $K$ is large enough, which may even show a non-monotone trajectory of the $\text{HR}_{\text{W}}$. 
However, modelling the $K$-point distribution with a large $K$ is difficult given that this involves $2(K-1)$ parameters for a latent distribution, especially if the cluster size is small.  
On the contrary, within the $\mathcal{AF}$ one is remunerated with $\lvert \Omega \rvert = \infty$ (except for the $\psi\mathcal{B}$ case)
which can be important for extreme observations where events occur very early in the lifespan. 
However, the parametric constraints of the $\mathcal{AF}$ on the trajectory of $\text{HR}_{\text{W}}$ should always be taken in consideration and challenged by the $K$-point distribution whenever possible. } 

This analysis can be extended to the across-stratum HR, $\text{HR}_\text{A}$, if there is a model for the distribution of individual heterogeneity  within the $\mathcal{AF}$.
For that purpose, we allow the parameters to depend on a stratum-specifying set of covariates $\tilde{\boldsymbol{x}}$, which determine the parameters of individual heterogeneity $\alpha(\tilde{\boldsymbol{x}})=\tilde{\boldsymbol{x}}^T  \boldsymbol{\zeta} ,\gamma(\tilde{\boldsymbol{x}}) = \exp\{\tilde{\boldsymbol{x}}^T\kappa\}$, with each parameter in the vectors $\boldsymbol{\zeta}, \boldsymbol{\kappa}$ being an element of $\mathbb{R}$.
For the sake of brevity, we denote $\alpha(\tilde{\boldsymbol{x}}_i),\gamma(\tilde{\boldsymbol{x}}_i),$ and $\mu(\tilde{\boldsymbol{x}}_i)$ by $\alpha_i ,\gamma_i,$ and $\mu_i$, respectively.
Figure~\ref{fig:HR_WA} shows the $\text{HR}_\text{A}$  for varying scenarios of $\alpha_i$ and $\alpha_{i'}$.
For $\alpha_i, \alpha_{i'} > 0$ (Figure~\ref{fig:HR_WA}\subref{fig:HR_app}), the $\text{HR}_\text{A}=\frac{\psi_i}{\psi_{i'}}$ for all $k\geq2$ and is undefined for $k=1$.
However, for $\alpha_i, \alpha_{i'} < 0$ (Figure~\ref{fig:HR_WA}\subref{fig:HR_ann}), the $\text{HR}_{\text{A}}(k)=\frac{\psi_i(\nu_i + k-1)}{\psi_{i'}(\nu_{i'}+k-1)}$, which approaches a constant ratio $\text{HR}_{\text{A}}(k)=\frac{\psi_i}{\psi_{i'}}$ for large $k$.
Note that the $\text{HR}_{\text{A}}$ might be greater or less than one for all $k$, but can also cross the threshold of one with increasing $k$. 
If $\tilde{\boldsymbol{x}}$ is for example an experimental treatment indicator (in a univariate context), the $\text{HR}_\text{A}$  represents a heterogeneous treatment effect which might identify sub-groups within the population for which the treatment is harmful. 
For $\alpha_i < 0, \alpha_{i'} > 0$ (Figure~\ref{fig:HR_WA}\subref{fig:HR_anp}), $\text{HR}_\text{A}(1) \equiv \infty$, 
\newnewnew{as for stratum $i'$ there is a latent sub-population that is not susceptible to the event of interest, whereas for stratum $i$ all latent sub-populations are susceptible. 
For $k \geq 2$ and $\text{HR}_\text{A}(k) = \frac{\psi_i(\nu_i + k-1)}{\psi_{i'}(k-1)}$. 
Another scenario, not explicitly shown in Figure \ref{fig:HR_app} and \ref{fig:HR_anp}, is that one or both strata ($i$ and $i'$) might have (different) upper bounds of frailty in the $\psi\mathcal{B}$ case.  
In such a case, the stratum with the larger upper bound (which could still be $\infty$), say $i'$, could be considered more vulnerable, since that stratum has a higher proportion of individuals who are expected to experience the event very early, namely those with a frailty value greater than $\psi_i b_i$. 
}

\newnew{Note that the parameters in the {formul\ae} of $\text{HR}_\text{A}$ and $\text{HR}_\text{W}$ (and hence the parameters as specified in the legends of Figure~\ref{fig:HR_WA}) do not uniquely identify the parameters of the frailty distribution, i.e. for a given trajectory of $\text{HR}_{A},\text{HR}_{W}$ there is an infinite set of $(\alpha,\gamma)$ or $(\alpha_i,\gamma_i)$ and $(\alpha_{i'},\gamma_{i'})$, respectively, that induce the same trajectory but with a different distributions of the RCs which did not need to be specified for Figure~\ref{fig:HR_WA}. 
This shows that the analysis of the frailty model has always two branches. 
The first branch is the analysis of HRs, which indicate the meaning of being in a particular latent RC (in a particular stratum) relative to another latent RC or to another observable stratum in the same latent RC. 
On the one hand, $\text{HR}_{\text{W}}$ can help to assess the importance of individual heterogeneity, e.g. if $\text{HR}_{\text{W}}(k)$ is large, then latent RC membership has a large effect on expected survival. 
On the other hand, comparing $\text{HR}_{\text{W}}$ across strata or analysing the $\text{HR}_{\text{A}}$ gives an account of random covariate effects where, e.g., covariates with a beneficial effect on survival or covariates with partly beneficial, partly detrimental effects might be detected. 
The second branch of the analysis is the distribution of RCs across strata, which may indicate differences in the distribution of risk-taking behaviour and predisposition across strata, e.g. by indicating a stratum with a heavier tail of vulnerable RCs. 
Taken together, the analysis of HRs and the distribution of RCs can provide thorough analytical explanations in terms of selection and random covariate effects that can help to explain the trajectories of population survival curves (where the RCs are marginalised out), i.e. explanations for why the survival curves of two strata come closer or even cross over time. }

%
 \label{Frailty}

\section{Estimation}

In this section, all time-dependent quantities are evaluated at the monitored (censored or uncensored) event times of the individuals.
We delete the argument from the expression and indicate the corresponding quantity with a subscript,
e.g. $\Lambda^{(j)}_{i} = \exp\{ {\boldsymbol{x}_i^{(j)}}^T\boldsymbol{\beta}^{(j)}\}\Lambda^{(j)}_0(t^{(j)}_i)$. 
Furthermore, let $A \in \mathbb{P}(\{1,\dots,J\})$, where $\mathbb{P}$ denotes the power set.
Then, $\Lambda^{(A)}_i = \sum_{j \in A} \Lambda_i^{(j)}$ and $\Lambda_i^{(-A)} = \sum_{j \notin A } \Lambda_i^{(j)}$.
Note that we define $\Lambda_i^{(\emptyset)} = 0$ and $\Lambda_i = \Lambda^{(1,\dots,J)}_i$.

We develop estimation routines for case I interval-censored data.
In the case of case I interval-censored data it is only known whether the event occurred during follow-up or not but the exact event time is unknown.
For multivariate cases ($J>1$) it is easier to understand the likelihood if one starts by exploiting the conditional independence assumption of $T_i^{(j)}$ and $T_i^{(j')}, j \neq j'$, given $Z_i=z$:

\begin{align}
	L(\boldsymbol{\theta},\boldsymbol{\lambda}_0, \boldsymbol{\beta};\text{data}) &= \prod_{i=1}^{n} \int_0^{\infty} \prod_{j = 1}^J (1-\exp\{-z\Lambda_{i}^{(j)}\})^{d_{i}^{(j)}} \exp\{-z\Lambda_{i}^{(j)}\}^{1-d_{i}^{(j)}}g_i(z)dz \nonumber  \\
&= \prod_{i=1}^{n} \int_0^{\infty} \sum_{A \in \mathbb{P}(d_{i})} (-1)^{|A|}\exp\{-z_i(\Lambda_i^{(A)} + \Lambda_i^{(-d_i)})\} g_i(z)dz \nonumber \\
&=  \prod_{i=1}^{n}\sum_{A \in \mathbb{P}(d_{i})} (-1)^{|A|} \mathcal{L}(\Lambda_i^{(A)} + \Lambda_i^{(-d_i)}). \label{Lcs}
\end{align}
where $d_i^{(j)}$ is the observational unit and target-specific event indicator (equal to one if the event occurred during the follow-up, zero otherwise), and $d_i$ is the set of targets on which the $i^{\text{th}}$ observational unit had an event. 
The vector $\boldsymbol{\lambda}_0$ contains all parameters of the baseline hazard rates, $\boldsymbol{\beta} = [\boldsymbol{\beta}^{(0)}, \boldsymbol{\beta}^{(1)},\dots, \boldsymbol{\beta}^{(J)}]$, and $\boldsymbol{\theta} = [\boldsymbol{\zeta},\boldsymbol{\kappa}]$.

Quasi-Newton optimization routines were applied for optimizing the corresponding log-likelihood based on \eqref{Lcs}. 
We choose BFGS as the standard method, as implemented in R version 4.2.2 \citep{R.Team.2014}. 
Standard errors (SE) are obtained via the Hessian of the log-likelihood, which is approximated by Richardson extrapolation as implemented by \cite{Gilbert.2019}. 
The delta method was applied where necessary to obtain SE: confidence intervals (CIs) are based on $\ln{}$ or $\ln\{ - \ln{} \}$  transformations if the parameter of interest is greater than zero or between zero and one respectively, and are then transformed back to the scale of interest. 

We provide algorithms that are able to fit univariate and multivariate frailty models for case I interval-censored data. 
The frailty distributions can be stratified by a (multi-level) factor.
The frailty distribution might either be $\mathcal{AF}$ or from the power variance family (both parameters can be estimated).
The baseline hazard can be chosen to be piecewise-constant or the parametric generalized gamma distribution \citep{Cox.2007} or one of its special cases, respectively.
\new{Covariates can be added in proportional hazards manner.
Overdispersion parameters might be added by means of the Dirichlet compound multinomial distribution.
Implementations are available on GitHub (https://github.com/time-to-MaBo/Addamsfamily/).}
%
 \label{Likelihood}

\section{Applications} \label{app}

We illustrate the $\mathcal{AF}$ in the context of multivariate case I interval-censored data on the human papillomavirus (HPV), obtained from a serological survey in the Netherlands (PIENTER-2);
see \cite{Mollema.2009} for details on PIENTER-2 and \cite{Scherpenisse.2012} for an investigation of the respective HPV dataset. 
The data were collected in the years 2006 and 2007 and cover people aged 0 to 79.
Participants were asked to complete a questionnaire and to provide a blood sample \citep{Mollema.2009}.
By means of the blood samples, the level of antibodies regarding the high-risk HPV types 16, 18, 31, 33, 45, 52, and 58 were determined in order to detect past infections. 
\newnewnew{Therefore, at the time of observation, it is only known whether the study participants have had an infection in the past or not, but it is never known exactly when the potential infection occurred, resulting in case I interval-censored data, also known as current status data \citep{Sun.2007}. }
Note that at the time of data collection the Dutch national immunization programme did not include a vaccine against HPV.

We analysed the nationwide sample including oversampled migrants and applied weighting factors to make the sample representative for the Dutch population. 
We excluded individuals in their first year of life from the analysis, as maternal antibodies could be transmitted to the infant transplacentally or through breastfeeding \citep{Rintala.2005}.
This left us with a sample size of $n=6384$ individuals\newnewnew{ aged 2 to 80}.  
The weighted proportion of females in the dataset is $49.9$\% (unweighted: $54.4$\%). 

The observed time is the individuals' age at the date of serological monitoring.
The event indicator $d_i^{(j)}=1$ means that individual $i$ is seropositive with respect to pathogen $j$, $j \in \{\text{HPV}16,\text{HPV}18,\text{HPV}31,\text{HPV}33,\text{HPV}45,\text{HPV}52,\text{HPV}58\}$, and seronegative and still susceptible if $d_i^{(j)}=0$.
Seroprevalence is interpreted as a proxy for past infections. 
Note, however, that there is a time-lag between the infection and the time of seroconversion, as well as a difference in the number of individuals who were infected with HPV and those who seroconverted:
in previous studies,  antibodies could not be detected for about 20-50\% of females who were carriers of HPV DNA.
However, antibody responses are relatively stable over time and hence, the study of the population's seroprevalence might yield important insights; see \cite{Scherpenisse.2012} for a discussion.

We consider the following models for the individual hazard rates,
$\lambda_i^{(j)}(t) = Z_i \lambda_0^{(\text{sex:}j)}(t)$, $\text{sex} \in \{m,f\}$, where the target-specific baseline hazard $\lambda_0^{(\text{sex:}j)}(t)$ is either sex-stratified (sex-stratified baseline hazard model), or non-stratified $\lambda_0^{(\text{sex:}j)}(t) = \lambda_0^{(j)}(t)$ (non-stratified baseline hazard model).  
\newnewnew{The purpose of stratifying baseline hazards by sex is twofold. 
The first is to investigate whether it is justified to estimate baseline hazards jointly for both sexes, and the second is to investigate whether a potential difference in the distribution is better explained by stratified baseline hazards than by different distributions of individual heterogeneity. }
In any case the target (and potentially sex) specific hazard rate is piecewise constant with a unique parameter within the intervals $[0;5)$, $[5;10)$, $[10;20)$, $[20;30)$, $[30;40)$, $[40;50)$, $[50;65)$, $[65;80)$.
The frailty $Z$ is either sex-stratified, i.e. $Z_i \sim \mathcal{AF}(\alpha_{\text{sex}_i},\gamma_{\text{sex}_i})$ (sex-stratified RE model), or non-stratified, i.e. $Z_i \sim \mathcal{AF}(\alpha,\gamma)$ (non-stratified RE model). 
Note that in both cases $\mu_m \equiv 1$, $\mu_f \in \mathbb{R}_{>0}$ except for the stratified-hazard model where $\mu_f$ is also set to one for the sake of identifiability. 
The stratified RE model might be better able to reflect differing patterns in individual heterogeneity due to biological and
environmental predisposition as well as a different distribution of risk-related behaviour across males and females.
We combine the stratification status of the baseline hazard with the stratification status of the RE.


An HPV infection can be transmitted via skin-to-skin contact, often - though not exclusively (see, e.g., \cite{Syrjanen.2010}, \cite{Rintala.2005} or \cite{Meyers.2014}) -  via sexual intercourse \citep{Gavillon.2010}. 
Therefore, individual – and typically unobserved – behaviour is an important determinant of an individual's risk of contracting HPV.
Frailty models have been used previously to incorporate unobservable individual heterogeneity in the transmission of infectious diseases (see, e.g., \cite{Unkel.2014} or  \cite{Hens.2009}).
Moreover, we suspect that there may be distinct jumps in the individual hazard rates due to differences in individual behaviour that are relevant for transmission, e.g. comparing individuals who have no sex at all to individuals who have (see, e.g., \cite{Richardson.2000}, \cite{Burchell.2006}), or whether the individuals use condoms or not (see, for example, \cite{Lam.2014} or \cite{Nielson.2010}). 
More formally, non-Gaussian and discrete omitted covariates may be the most important drivers of individual heterogeneity, and thus a discrete frailty model may be particularly appropriate here.

We start with a bivariate analysis including HPV16 and HPV18. An
extension to higher variate data with data on seven types of HPV
follows.

\hypertarget{bivariate-data-analysis}{%
\subsection{Bivariate Data Analysis}\label{bivariate-data-analysis}}

\begin{figure}[!ht]
  \centering
\begin{subfigure}{1\textwidth}
\includegraphics[width=.94\linewidth]{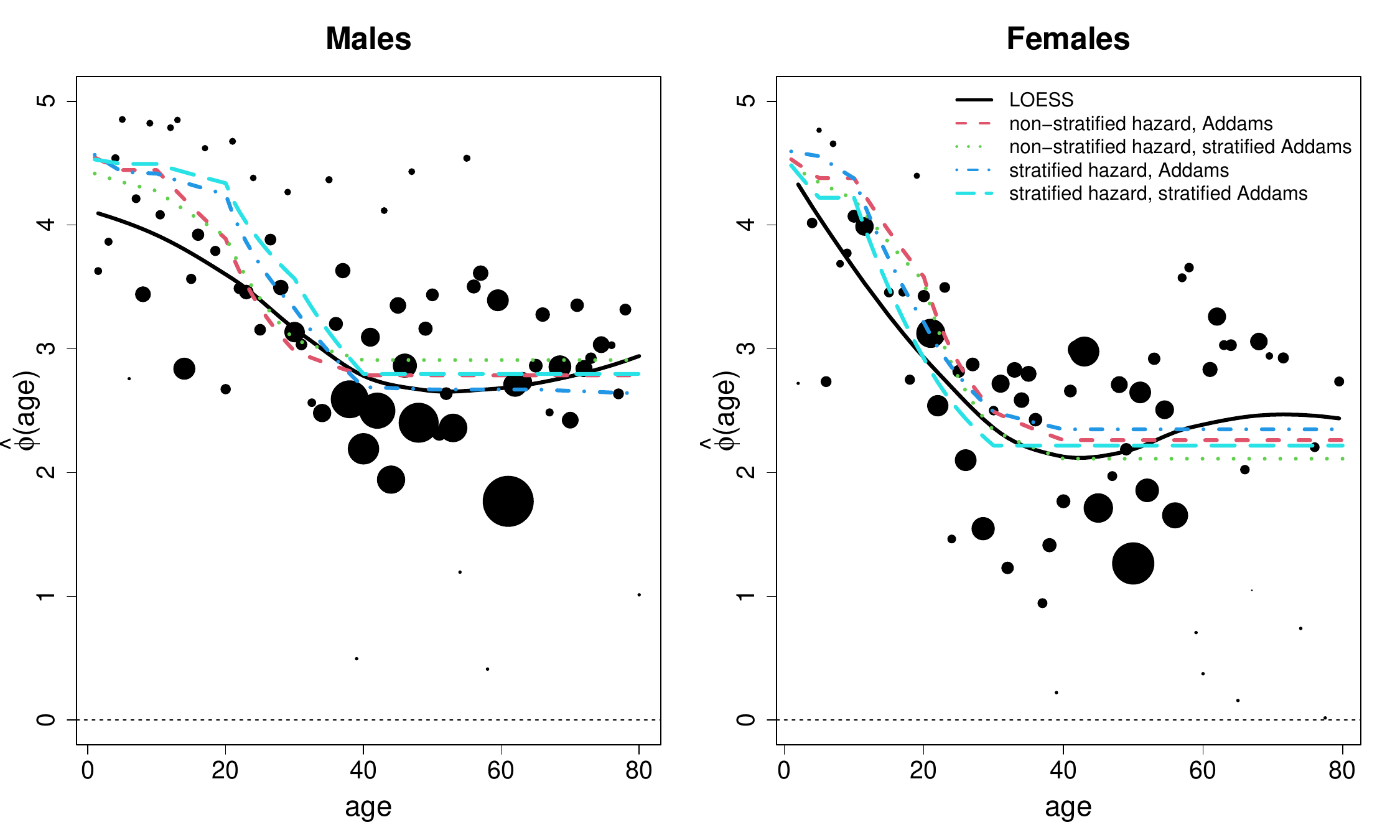}
   \caption{Observed association between HPV16 and HPV18 in terms of a $\phi$-plot. Black dots refer to cohort- and sex-specific non-parametric estimates, size proportional to precision. The black solid line is the corresponding LOESS. Other dotted and dashed lines are estimates resulting from corresponding parametric model.}
   \label{fig:Phi-Plot}
\end{subfigure}

   \begin{subfigure}{1\textwidth}
    \includegraphics[width=.94\linewidth]{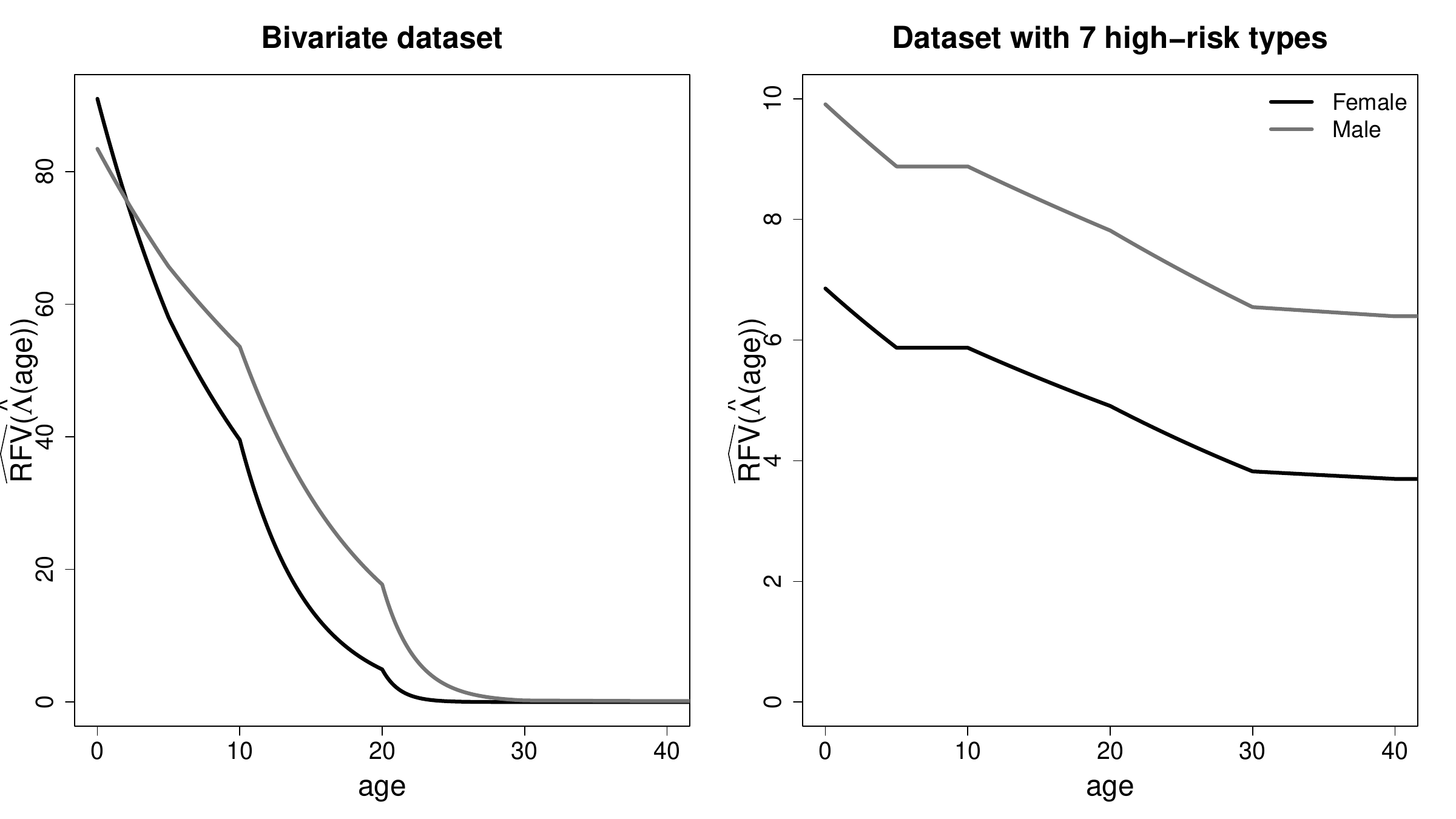}
   \caption{\newnew{RFV of bivariate (left) and higher variate (right) dataset of non-stratified hazard, stratified RE model. Note the different scales on the y-axis. The curves reach a plateau at around 30 years of age, hence, the x-axis was cut-off after 40 years. Note that also the seroprevalence curves ($1-P(T^{(j)}>t)$) reach a plateau around this time (not shown).}}
   \label{fig:RFV7vs2}
\end{subfigure}
\caption{\new{Observed association between HPV types in the PIENTER-2 data.}}
   \end{figure}

We begin by exploiting the nested structure of the models for model
selection. The stratification of the RE is statistically significant on
conventional levels by means of a likelihood ratio test (LRT) no matter
the stratification status of the baseline hazard. The null-hypothesis is
\(H_0\): \(Z_{m} \sim Z_{f} \sim \mathcal{AF}(\alpha,\gamma)\)
vs.~\(H_1: Z_{\text{sex}} \sim \mathcal{AF}(\alpha_{\text{sex}},\gamma_{\text{sex}})\).
Note that the expectation parameter \(\mu_f\) is not included in the
hypothesis. In the case of non-stratified baseline hazards the LRT test
statistic equals 29.700 on 2 degrees of freedom (p-value \(\approx\) 0).
In the case of sex-stratified baseline hazards the LRT test statistic
equals 30.822 on 2 degrees of freedom (p-value \(\approx\) 0). Better
performance of stratified RE models is also suggested by the
\(\phi\)-plot which can be seen in
Figure~\ref{fig:Phi-Plot}. The measure \(\phi\) is an
association measure for bivariate current status data introduced by
\cite{Unkel.2012}, \(\phi>0\) indicates positive and \(\phi<0\) negative
association. Additionally, \(\phi\) tracks
\(\ln\{1+\operatorname{RFV}(\boldsymbol{t})\}\) with a time-lag. It can
be observed that the association between HPV16 and HPV18 is higher for
females early in life, but declines more strongly than for males. This
is likely to be the reason for the success of stratified RE models here
as those models are able to choose a distinct intercept and slope of the
RFV across the sexes. We choose the sex-stratified RE model for further
analysis.

In terms of AIC, the non-stratified baseline hazard model
\(\lambda^{(j)}_0(t)\), \(j \in \{\text{HPV} 16, \text{HPV} 18\}\)
performs better than sex-stratified-baseline hazard model
\(\lambda_0^{(\text{sex:}j)}(t)\) (9647 vs.~9656). Thus, we choose the
non-stratified baseline hazard, stratified RE model for further
analysis.

A LRT for a constant RFV (CRF),
i.e.~\(H_0: Z_{\text{sex}} \sim \mathcal{G}(\gamma_{\text{sex}})\)
vs.~\(H_1: Z_{\text{sex}} \sim \mathcal{AF}(\alpha_{\text{sex}},\gamma_{\text{sex}})\)
for males and females, yields a test-statistic of 93.514 on 2 degrees of
freedom (p-value \(\approx 0\)). Hence, the hypothesis on a constant RFV
(CRF) can also be rejected. A constant pattern of the RFV is also not
suggested by Figure~\ref{fig:Phi-Plot}, where it can be
observed that association is constantly falling up to the age of around
\(50\). Considering all tests and pairwise AIC comparisons we choose the
non-stratified baseline hazard, stratified \(\mathcal{AF}\) model for
final analysis.

\begin{table}[!htb]
        \centering
      \caption{\newnewnew{Estimated RFV parameters (above dashed line) and resulting estimated frailty distribution parameters (below dashed line) for non-stratified hazard, stratified RE model.} Parentheses below point estimates show $95$\%-CIs.}
       
\begin{tabular}{lcc}
\toprule
  & male & female\\
\midrule
$ \hat{\alpha}_{\text{sex}} $ & $\underset{(-0.809;-0.196)}{-0.502}$ & $\underset{(-5.008;-0.757)}{-2.882}$\\
$ \hat{\gamma}_{\text{sex}} $ & $\underset{(66.629;104.509)}{83.447}$ & $\underset{(60.167;137.621)}{90.996}$\\
\hdashline
$ \hat{\psi}_{\text{sex}} $ & $\underset{(0.243;1.038)}{0.502}$ & $\underset{(0.336;2.66)}{0.946}$\\
$ \hat{\nu}_{\text{sex}} $ & $\underset{(0.009;0.016)}{0.012}$ & $\underset{(0.006;0.018)}{0.011}$\\
$ \hat{\pi}_{\text{sex}} $ & $\underset{(0.002;0.015)}{0.006}$ & $\underset{(0.009;0.11)}{0.031}$\\
\bottomrule
\end{tabular} \label{tab:RFVpar} \\ 
  \end{table}

\begin{table}[!htb]
        \centering
      \caption{Estimated distribution of RCs and across stratum analysis for stratified RE, non-stratified hazard model. Parentheses below point estimate show 95\%-CI.}
       
\begin{tabular}{lllllll}
\toprule
\multicolumn{1}{c}{ } &  \multicolumn{2}{c}{$\hat{P}\big( {Z}_{\text{sex}} \leq \hat{z}_{\text{sex},(k)} \big)$} & \multicolumn{1}{c}{ } & \multicolumn{2}{c}{$\hat{z}_{\text{sex},(\text{k})}$} &  \multicolumn{1}{c}{ }  \\
 \cmidrule(l{3pt}r{3pt}){2-3} \cmidrule(l{3pt}r{3pt}){5-6}
$k^{\text{th}}$ RC  & males & females & $\frac{\hat{P}\big( {Z}_{f} \leq {z}_{f,(k)} \big)}{\hat{P}\big( {Z}_{m} \leq {z}_{m,(k)} \big)}$ & males & females  & $\widehat{\text{HR}}_{\text{A}}(k)$\\
\midrule
$1^{\text{st}}$ &  $\underset{(0.932;0.949)}{0.941}$ & $\underset{(0.956;0.97)}{0.964}$ & $\underset{(1.013;1.036)}{1.024}$ & $\underset{(0.002;0.015)}{0.006}$ & $\underset{(0.004;0.026)}{0.01}$ & $\underset{(1.411;2.01)}{1.684}$\\
$2^{\text{nd}}$ &  $\underset{(0.949;0.955)}{0.952}$ & $\underset{(0.971;0.976)}{0.974}$ & $\underset{(1.013;1.032)}{1.023}$ & $\underset{(0.245;1.053)}{0.508}$ & $\underset{(0.341;2.684)}{0.956}$ & $\underset{(1.064;3.325)}{1.881}$\\
$3^{\text{rd}}$ & $\underset{(0.956;0.96)}{0.958}$ & $\underset{(0.977;0.98)}{0.978}$ & $\underset{(1.017;1.027)}{1.022}$ & $\underset{(0.489;2.091)}{1.011}$ & $\underset{(0.677;5.344)}{1.902}$ & $\underset{(1.061;3.336)}{1.882}$\\
$4^{\text{th}}$ & $\underset{(0.96;0.963)}{0.961}$ & $\underset{(0.98;0.983)}{0.982}$ & $\underset{(1.017;1.025)}{1.021}$ & $\underset{(0.732;3.13)}{1.513}$ & $\underset{(1.014;8.004)}{2.848}$ & $\underset{(1.061;3.34)}{1.882}$\\
$5^{\text{th}}$ & $\underset{(0.963;0.966)}{0.964}$ & $\underset{(0.983;0.985)}{0.984}$ & $\underset{(1.017;1.024)}{1.02}$ & $\underset{(0.975;4.168)}{2.016}$ & $\underset{(1.35;10.664)}{3.794}$ & $\underset{(1.06;3.341)}{1.882}$\\
\bottomrule
\end{tabular} \label{tab:Pz} \\ 
  \end{table}

The RFV parameter estimates for the stratified RE, non-stratified
baseline hazard models can be seen in
Table~\ref{tab:RFVpar}. The estimated RFV (CRF) is
decreasing for both sexes. The estimated RFV parameters indicate higher
heterogeneity across clusters or association within a cluster for
females early on, as indicated by the intercept of the RFV
(\(\hat{\gamma}_f > \hat{\gamma}_m\)). However, the descent is more
strongly for females
(\(\lvert \hat{\alpha}_f \rvert > \lvert\hat{\alpha}_m \rvert\)) and
consequently it is estimated that heterogeneity/association is stronger
for males from the 4\(^\text{th}\) year of life onwards; \newnew{see left-hand panel of Figure \ref{fig:RFV7vs2}. These results}
are also supported by the non-parametric estimates of \(\phi\) in
Figure~\ref{fig:Phi-Plot}.

The estimated distribution corresponds to a
\(\hat{\psi}_{\text{sex}} \mathcal{NB}_{>0}(\hat{\nu}_{\text{sex}},\hat{\pi}_{\text{sex}})\)
for males and females. The mean parameter
\(\hat{\mu}_f =\)
\({0.328}\ (95\%\text{-CI } [0.091;1.182])\)and is highly insignificant as
indicated by the \(95\%\)-CI. The resulting distribution parameters can also 
be found in Table~\ref{tab:RFVpar}. The estimated mean
\(\hat{\mu}_{f}\) indicates lower expected frailty (and therefore lower
population hazard) for females initially. However, the mean parameter
has to be interpreted in the context of its distribution. Let
\(\mu_{\text{sex}}(\boldsymbol{t}) = \operatorname{E}(Z|\boldsymbol{T}>\boldsymbol{t},\text{sex})\).
The limit of the conditional expectation of the frailty is
\(\mu_{\text{sex}}([{\infty},{\infty}]) = \psi_{\text{sex}}\nu_{\text{sex}}\).
With the estimates from Table~\ref{tab:RFVpar},
\(\hat{\mu}_{f}([{\infty},{\infty}]) =\) 0.01
\(> \hat{\mu}_{m}([{\infty},{\infty}]) =\) 0.006 follows and the initial
order of the expectations is reversed. In this example, this leads to
the estimated population seroprevalence
\(\hat{P}(T^{(\text{HPV} 16)} \leq \text{age})\) being higher for males
early in life but from 12 years of life onwards, females start to catch
up and finally cross the curve of males at 25 years of life (not shown).
We will discuss the reason for the switching order of
\(\hat{\mu}_f(\boldsymbol{t})\), and \(\hat{\mu}_m(\boldsymbol{t})\)
that finally leads to crossing seroprevalence curves by analysing the
distribution of the frailties in the paragraphs below.

Table~\ref{tab:Pz} shows an excerpt of the distribution
of the RCs. We interpret the distribution of the RCs as the distribution
of stratum-relative risk-related behaviour and predisposition. The bulk
of the population is estimated to be in the lowest RC, though there is
more probability to the right of the lowest RC for males. The ratio of
the cumulative probabilities between females and males is always above
one, also indicating a more heavy tail for males. The heavier tale of
the distribution of latent RCs for males is the reason for
\(\mu_m>\hat{\mu}_f\).

The numerical value of the frailties then assigns a magnitude related
interpretation to the distinct RCs. The estimated support shows that
females have a higher category-related hazard in each RC (see the last column
Table~\ref{tab:Pz}). Across strata, given the
\(k^{\text{th}}\) RC, the conditional or RC-related HR,
\(\widehat{\text{HR}}_{\text{A}}(k) = \frac{\hat{z}_{f,(k)}\hat{\lambda}^{(j)}_0(t)}{\hat{z}_{m,(k)}\hat{\lambda}^{(j)}_0(t)}\),
is 1.684 in the important first category. In this case, the
\(\widehat{\text{HR}}_{\text{A}}(k)\) approaches its limit (with respect
to \(k\)), \(\frac{\hat{\psi}_{f}}{\hat{\psi}_{m}} =\) 1.883, fast due
to small values of \(\hat{\nu}_{\text{sex}}\). Higher RC-related hazard
for females is the reason for \(\hat{\mu}_f(\boldsymbol{t})\) surpassing
\(\hat{\mu}_m(\boldsymbol{t})\) with time progressing: the individuals
belonging to the tale of the distribution of the RCs start to
seroconvert early. This selection effect is more pronounced for males
due to the heavier tale of the distribution of RCs. Due to extreme
individuals within the male population seroconverting more quickly, the
higher RC-related hazard for females causes
\(\hat{\mu}_f([t,0])>\hat{\mu}_m([t,0])\) from 12 years of life onwards.

Within stratum,
\({\widehat{\text{HR}}_{\text{W}}}(1)=\)
\({94.878}\ (95\%\text{-CI } [60.448;148.919]) \) for females and
\({84.949}\ (95\%\text{-CI } [65.475;110.215]) \) for males, i.e. being in the
second instead of the lowest RC is estimated to be more hazardous for
females than for males even from a relative perspective. The
\(\widehat{\text{HR}}_{\text{W}}(k)\) then approaches its limit
\(\frac{k}{k-1}\) immediately because \(\hat{\nu}_{\text{sex}}\) is
small for males and females.

Differences in unobserved heterogeneity across the sexes are reflected
by the support and the distribution of the RCs. Given that HPV is a
sexually transmitted disease, the membership to a certain RC is partly
governed by stratum-relative (sexual) behaviour in that sense, that
having, for example, a higher number of sexual partners than some
stratum reference should put one in a higher RC than the reference
individual. It is tempting to interpret the difference in magnitude of a
given RC on the conditional hazard rate across the sexes. For the human
immunodeficiency viruses, for example, it is known that male to female
transmission is more likely than female to male transmission (see
\cite{Nicolosi.1994} or
\cite{EuropeanStudyGrouponHeterosexualTransmissionofHIV.1992}). Assuming
that each RC comprises the same set of sexual behaviour across the
sexes, \(z_{f,(k)}>z_{m,(k)}\) for all \(k\), could also hint on a
higher susceptibility of females with respect to an infection with HPV16
and HPV18 per relevant contact. However, the RCs are anchored in the
stratum and do not necessarily imply the same behaviour across the
sexes. 
Hence, this interpretation is highly speculative and assumption based.

\hypertarget{higher-variate-analysis}{%
\subsection{Higher Variate Analysis}\label{higher-variate-analysis}}

When including all seven high-risk types of HPV for which we have data,
the direction of interpretation is largely similar to that of the
bivariate case, and mainly the magnitude changes. The estimated RFV
parameters are
\(\hat{\alpha}_m =\) \({-1.359}\ (95\%\text{-CI } [-1.8;-0.918])\),
\(\hat{\gamma}_m =\) \({9.908}\ (95\%\text{-CI } [8.928;10.995])\),
\(\hat{\alpha}_f =\) \({-2.005}\ (95\%\text{-CI } [-2.5;-1.509])\),
\(\hat{\gamma}_f =\) \({6.855}\ (95\%\text{-CI } [6.143;7.649])\). The
heterogeneity/association is less extreme than in the bivariate case
early in life. However, the association remains at larger levels
compared to the bivariate case, as shown in Figure~\ref{fig:RFV7vs2},
indicating that association remains high throughout life. 
It can also be seen that the RFV (CRF) of
females is always below that of males, indicating greater heterogeneity
due to individual factors for males throughout the entire time period.
As the level of association differs strongly between the bivariate case
above and the higher variate case with seven high-risk HPV types, a
shared frailty model might be seen as inadequate to capture the patterns
of association between the various types of HPV. Therefore, a correlated
frailty model may be more appropriate. The shared frailty model might be
chosen for its simplicity, however, if the specific types of HPV are
less relevant to the research question but, for example, the prognostic
factor of one ``anonymous'' high-risk type on another ``anonymous''
high-risk type is investigated.

The (initial) expectation of the frailty is virtually identical for
males and females; \({\hat{\mu}_f} =\)
\({0.955}\ (95\%\text{-CI }[0.883;1.033])\). In the higher variate case the
distribution of the RCs is not as much focused on the lower categories.
The tail is again more heavy for males (not shown). The
\(\text{HR}_{\text{A}}(k)>1\) for all \(k\) again indicates higher
RC-related hazard for females. The
\({\widehat{\text{HR}}_{\text{W}}(1)}\) is
less extreme in the higher variate case than in the bivariate case:
\({9.86}\ (95\%\text{-CI }[8.627;11.268])\) for females and
\({12.267}\ (95\%\text{-CI }[10.786;13.95])\) for males. Note that the order of
\(\widehat{\text{HR}}_{\text{W}}(1)\) for males and females changes when
comparing this to the bivariate scenario.

\section{Conclusion} \label{conc}

In this paper, we discuss the Addams family of discrete frailty distributions, which has been \newnewnewnew{conceptualised} by \cite{Farrington.2012} for modelling individual heterogeneity in time-to-event models. 
We further examine the properties of the conditional time-to-event model induced by the Addams family and develop estimation routines for multivariate case I interval-censored data.

For discrete frailty distributions, the RFV (CRF) approaches either infinity or zero (one) over the course of time, where the distinction is made by the minimum of the support of the frailty being zero or greater than zero respectively.
Few discrete frailty distributions are able to manipulate the support via its parameters to choose the long-term behaviour of the two functions accordingly, but this typically involves the edge of the parameter space; 
see \cite{Bardo.2023a} for a discussion.
For the Addams family of discrete frailty distributions, the minimum of the support can either be zero, resulting in a cure rate model, or greater than zero without involving the edge of the parameter space.
Consequently, the RFV (CRF) is either monotonically increasing or monotonically decreasing, again without involving the edge of the parameter space.
Through the introduction of a scaling parameter, the Addams family is also able to increase or flatten the slope of the RFV (CRF) and might even approach a constant by approaching its continuous exception, a shape that is impossible for discrete shared frailty model to generate.
This makes the Addams family a useful general-purpose modelling approach.

A unique feature of the Addams family is that the support of the discrete frailty distribution varies with its parameters and is hence subject to estimation.
We suggest interpreting the support as ordered latent risk categories.
This feature allows for a unique analysis of the latent model through calculating hazard ratios across different latent risk categories.
We call this the within-stratum hazard ratio.
If a model is imposed on the distribution parameters of the Addams family, this analysis can be enriched by the across-stratum hazard ratio,
i.e. the hazard ratio of a given latent risk category for different strata that are defined by covariates.
This type of analysis could also be performed with the discrete $K$-point distribution.
However, there is no counterpart to this analysis for continuous frailty distributions, or for discrete frailty distributions for which the support is fixed.
Thus, the Addams family and the $K$-point distribution offer the possibility to analyse the latent model, which may include heterogeneous covariate effects, more thoroughly.

The analysis of the latent model via the within-stratum hazard ratio might help to understand the importance of individual heterogeneity.
In that sense, individual heterogeneity might be regarded as important if the within-stratum hazard ratios are large and vice versa.
The analysis of the across-stratum hazard ratio may reveal structural differences in individual heterogeneity across covariates, prompting a discussion of the reasons for this.
Furthermore, it is an advantage that this analysis can be carried out via the analysis of time-invariant hazard ratios, which is arguably the most common way to analyse data in time-to-event analysis.
Hence, we think the interpretation of the within-stratum and across-stratum hazard ratios is a particular suitable way to communicate with an audience outside of statistics.
Due to the discrete latent random variables, the analysis can further be enriched with probabilities of belonging to a certain risk category (conditional on survival up to $t$) and the corresponding differences across strata, which should be straightforward to communicate.
We applied the Addams family to multivariate case I interval-censored infection data and allowed the distribution of individual heterogeneity to differ for males and females. 
Males are found to have a higher probability for more hazardous categories, possibly reflecting a more cautious behaviour in the female population compared to males. 
However, the estimated hazard in each risk category is higher for females than for males, which might reflect a higher biological burden with respect to the susceptibility of HPV. 
There was no evidence for the existence of a non-susceptible sub-group, neither in the bivariate data set, including HPV 16 and HPV 18, nor in the data set containing seven high-risk types of HPV. 







\section*{Competing interests}
No competing interest is declared.



\section*{Acknowledgments}
We are grateful to Fiona van der Klis and Liesbeth Mollema from the Dutch National Institute for Public Health and the Environment (RIVM), for giving us permission to use the data from the second PIENTER study. 
The authors also thank the anonymous reviewers for their valuable suggestions. 
This work was supported by
the Deutsche Forschungsgemeinschaft
(DFG; German Research Foundation)	
[grant number UN 400/2-1].









\end{document}